\documentclass[11pt]{article}
\usepackage[superscript]{cite}
\usepackage{wordlike}
\usepackage{setspace}

\usepackage{hyperref}
\hypersetup{colorlinks,urlcolor=blue}
\usepackage{float}
\usepackage{pdflscape}
\usepackage{textcomp}
\usepackage{cite}
\usepackage{times}
\usepackage{latexsym}
\usepackage{hyperref}
\usepackage{booktabs}
\usepackage{amsmath}
\usepackage{threeparttable}
\usepackage{tabularx}
\usepackage{graphicx}
\usepackage{multirow}
\usepackage{longtable}
\usepackage{xcolor}
\usepackage{threeparttable}
\makeatletter
\definecolor{myblue1}{RGB}{47,84,150}
\newcommand{\printfnsymbol}[1]{%
  \textsuperscript{\@fnsymbol{#1}}%
}
\renewcommand{\section}{\@startsection%
  {section}%
  {0}%
  {0em}%
  {-\baselineskip}%
  {0.5\baselineskip}%
  {\color{myblue1}\Large\sffamily}}%
\renewcommand{\subsection}{\@startsection%
  {subsection}%
  {1}%
  {0em}%
  {-\baselineskip}%
  {0.5\baselineskip}%
  {\color{myblue1}\large\sffamily}}%
\makeatother

\renewcommand{\frame}{}

\title{\singlespacing A deep learning approach for automated detection of geographic atrophy from color fundus photographs}

\author{
\small
Tiarnan D. Keenan, BM BCh, PhD$^{1,*}$, Shazia Dharssi$^{1, 2,*}$, Yifan Peng, PhD$^{2,*}$, Qingyu Chen, PhD$^{2}$, Elvira Agr\'{o}n, MA$^{1}$, Wai T. Wong, MD$^{1,3}$, Zhiyong Lu, PhD$^{2}$, Emily Y. Chew, MD$^{1}$\\
1. National Eye Institute (NEI), National Institutes of Health (NIH), Bethesda, Maryland, United States;\\
2. National Center for Biotechnology Information (NCBI), National Library of Medicine (NLM), National Institutes of Health (NIH), Bethesda, Maryland, United States;\\
3. Unit on Microglia, National Eye Institute (NEI), National Institutes of Health, Bethesda, MD, USA;\\
* These authors contributed equally to this work.\\
}

\date{}

\begin{document}

\maketitle

\parindent=0em
\textbf{Taxonomy topics (2-6)}

deep learning; geographic atrophy; central geographic atrophy; Age-Related Eye Disease Study (AREDS); artificial intelligence (AI)
\vspace{1em}

\textbf{Corresponding Author(s)}

Emily Y. Chew, MD, National Eye Institute (NEI), National Institutes of Health (NIH), 9000 Rockville Pike, Bethesda, MD 20894, \url{echew@nei.nih.gov}, 301.496.6583

Zhiyong Lu, PhD, National Center for Biotechnology Information (NCBI), National Library of Medicine (NLM), National Institutes of Health (NIH), 8600 Rockville Pike, Bethesda, MD 20894, \url{zhiyong.lu@nih.gov}, 301.594.7089
\vspace{1em}

\textbf{Financial Support}: 
Supported by the intramural program funds and contracts from the National Center for Biotechnology Information/National Library of Medicine/National Institutes of Health, the National Eye Institute/National Institutes of Health, Department of Health and Human Services, Bethesda Maryland (contract HHS-N-260-2005-00007-C; ADB contract NO1-EY-5-0007). Funds were generously contributed to these contracts by the following NIH institutes: Office of Dietary Supplements, National Center for Complementary and Alternative Medicine; National Institute on Aging; National Heart, Lung, and Blood Institute, and National Institute of Neurological Disorders and Stroke. The sponsor and funding organization participated in the design and conduct of the study; data collection, management, analysis and interpretation; and the preparation, review and approval of the manuscript.
\vspace{1em}

\pagebreak
\parindent=2em

\section*{Abstract}
\vspace{-1.5em}
\hspace*{2em}\textbf{Purpose}: To assess the utility of deep learning in the detection of geographic atrophy (GA) from color fundus photographs; secondary aim to explore potential utility in detecting central GA (CGA).

\textbf{Design}: A deep learning model was developed to detect the presence of GA in color fundus photographs, and two additional models to detect CGA in different scenarios.

\textbf{Participants}: 59,812 color fundus photographs from longitudinal follow up of 4,582 participants in the Age-Related Eye Disease Study (AREDS) dataset. Gold standard labels were from human expert reading center graders using a standardized protocol.

\textbf{Methods}: A deep learning model was trained to use color fundus photographs to predict GA presence from a population of eyes with no AMD to advanced AMD. A second model was trained to predict CGA presence from the same population. A third model was trained to predict CGA presence from the subset of eyes with GA. For training and testing, 5-fold cross-validation was employed. For comparison with human clinician performance, model performance was compared with that of 88 retinal specialists.

\textbf{Main Outcome Measures}: Area under the curve (AUC), accuracy, sensitivity, specificity, precision.

\textbf{Results}: The deep learning models (GA detection, CGA detection from all eyes, and centrality detection from GA eyes) had AUC of 0.933-0.976, 0.939-0.976, and 0.827-0.888, respectively. The GA detection model had accuracy, sensitivity, specificity, and precision of 0.965 (95\% CI 0.959-0.971), 0.692 (0.560-0.825), 0.978 (0.970-0.985), and 0.584 (0.491-0.676), respectively, compared to 0.975 (0.971-0.980), 0.588 (0.468-0.707), 0.982 (0.978-0.985), and 0.368 (0.230-0.505) for the retinal specialists. The CGA detection model had equivalent values of 0.966 (0.957-0.975), 0.763 (0.641-0.885), 0.971 (0.960-0.982), and 0.394 (0.341-0.448), compared to 0.990 (0.987-0.993), 0.448 (0.255-0.641), 0.993 (0.989-0.996), and 0.296 (0.115-0.477). The centrality detection model had equivalent values of 0.762 (0.725-0.799), 0.782 (0.618-0.945), 0.729 (0.543-0.916), and 0.799 (0.710-0.888), compared to 0.735 (0.445-1), 0.878 (0.722-1), 0.703 (0.332-1), and 0.626 (0.273-0.979).

\textbf{Conclusions}: A deep learning model demonstrated high accuracy for the automated detection of GA. The AUC was non-inferior to that of human retinal specialists. Deep learning approaches may also be applied to the identification of CGA. The code and pretrained models are publicly available at \url{https://github.com/ncbi-nlp/DeepSeeNet}.

\pagebreak

Age-related macular degeneration (AMD) is the leading cause of irreversible vision loss in individuals older than 60 years of age in Western countries, and accounts for approximately 9\% of blindness worldwide\cite{congdon2004causes,wong2014global,quartilho2016leading}. Late AMD takes two forms, neovascular and atrophic, though these may coexist in the same eye. Geographic atrophy (GA) is the defining lesion of the atrophic form of late AMD. GA in AMD has been estimated to affect over 5 million people worldwide\cite{wong2014global,rudnicka2012age}. Unlike for neovascular AMD, no drug therapies are available to prevent GA, slow its enlargement, or restore lost vision; this makes it an important research priority\cite{keenan2018progression,rosenfeld2018preventing}. Rapid and accurate identification of eyes with GA could lead to improved recruitment of eligible patients for future clinical trials and eventually to early identification of appropriate patients for proven treatments.

Since the original description of GA by Gass\cite{gass1973drusen}, clinical definitions have varied between research groups\cite{schmitz-valckenberg2016geographic}. In the Age-Related Eye Disease Study (AREDS), it was defined as a sharply demarcated, usually circular zone of partial or complete depigmentation of the retinal pigment epithelium (RPE), typically with exposure of underlying large choroidal blood vessels, at least as large as grading circle I-1 (1/8 disc diameter in diameter)\cite{group2001randomized}. Sensitivity of the retina to light stimuli is markedly decreased (i.e., dense scotomata) in areas affected by GA. The natural history of the disease involves progressive enlargement of GA lesions over time, with visual acuity decreasing markedly as the central macula becomes involved\cite{keenan2018progression}.

The identification of GA by ophthalmologists conducting dilated fundus examinations is sometimes challenging. This may be particularly true for cases with early GA, comprising smaller lesions with less extensive RPE atrophy. In addition, the increasing prevalence of GA (through aging populations in many countries) will translate to greater demand for retinal services. As such, deep learning approaches involving retinal images, obtained perhaps using telemedicine-based devices, might support GA detection and diagnosis. However, these approaches would require the establishment of evidence-based and `explainable' systems that have undergone extensive validation and demonstrated performance metrics that are at least non-inferior to those of clinical ophthalmologists in routine practice.

Recent studies have demonstrated the utility of deep learning in the field of ophthalmology\cite{burlina2017automated,lam2018retinal,gargeya2017automated,gulshan2016development,wong2016artificial,venhuizen2018deep,asaoka2016detecting,peng2018deepseenet,chen2019multi}. Deep learning is a branch of machine learning that allows computers to learn by example. As applied to image analysis, deep learning enables computers to perform classification directly (unsupervised) from images, rather than through the recognition of features prespecified by human experts. This is achieved by training algorithmic models on images with accompanying labels (e.g. color fundus photographs categorized manually for the presence or absence of GA), such that these models can then be used to classify new images with similar labels. The models are neural networks that are constructed of an input layer (which receives, for example, the color fundus photograph), followed by multiple layers of non-linear transformations, to produce an output (e.g. GA present or absent). One recent deep learning model, DeepSeeNet\cite{peng2018deepseenet}, performed patient-based AMD severity classification with a level of accuracy higher than a group of human retinal specialists. Indeed, it also did this in an `explainable' way, by simulating the human grading process, which may improve levels of acceptability to ophthalmologists and patients. However, to the best of our knowledge, few studies have focused specifically on GA\cite{hu2013segmentation,feeny2015automated,ramsey2014automated,treder2018automated}, and the majority of these have concentrated on image segmentation tasks rather than automated detection of GA.

The primary aim of this study was to assess the utility of deep learning for the detection of GA from color fundus images. We also conducted experiments to assess the potential utility of deep learning for the detection of central GA (CGA) from color fundus images.

\section*{Methods}

\subsection*{Dataset}

The dataset used for this study (training and testing) was from the AREDS. The AREDS was a multi-center, prospective cohort study of the clinical course of AMD (and age-related cataract), as well as a phase III randomized clinical trial of nutritional supplementation for treatment of AMD and cataract\cite{age-relatedeyediseasestudyresearch1999age}. Its primary outcome was the development of advanced AMD, defined as CGA or neovascular AMD. The study design has been described previously\cite{age-relatedeyediseasestudyresearch1999age}. In short, 4,757 participants aged 55 to 80 years were recruited between 1992 and 1998 at 11 retinal specialty clinics in the USA. Based on color fundus photographs, best-corrected visual acuity, and ophthalmologic evaluations, participants were enrolled into one of several AMD categories. Institutional review board approval was obtained at each clinical site and written informed consent for the research was obtained from all study participants. The research was conducted under the Declaration of Helsinki.

As described previously, at baseline and annual study visits, comprehensive eye examinations were performed by certified study personnel using standardized protocols. At the baseline, 2-year, and annual study visits thereafter, stereoscopic color fundus photographs were taken of both eyes (field 2, i.e., 30\textdegree~imaging field centered at the fovea). Because of the inherent redundancy in a pair of stereoscopic photographs, for each eye, only one of the pair of photographs was used in the current study. In general, the left image of the pair was used, unless it was missing from the database, in which case the right image was used instead ($\sim$ 0.5\%). As a result, a total of 59,812 color fundus images from 4,582 participants were extracted from the AREDS dataset (Table~\ref{tab:Characteristics}).
\begin{table}[ht]
\centering
\small
\caption{Characteristics of images and participants in each partition according to the presence of geographic atrophy and central geographic atrophy.}
\label{tab:Characteristics}
\begin{tabular}{l@{\hspace{1ex}}l@{\hspace{1ex}}c@{\hspace{1ex}}c@{\hspace{1ex}}c@{\hspace{1ex}}c@{\hspace{1ex}}c}
    \toprule
 & \textbf{Total} & \multicolumn{5}{c}{\textbf{Partition}}\\
 \cmidrule{3-7}
 &  & 1 & 2 & 3 & 4 & 5\\
 \midrule
Images & 59,812 & 11,933 & 11,814 & 11,882 & 12,056 & 12,127\\
Participants & 4,582 & 924 & 907 & 915 & 919 & 917\\
Age: Mean, years (SD) & 69.4 (5.1) & 69.5 (5.1) & 69.4 (5.2) & 69.4 (5.2) & 69.6 (5.0) & 69.2 (5.0)\\
Gender (\% male) & 44.3\% & 44.9\% & 44.4\% & 45.2\% & 44.7\% & 42.2\%\\
Geographic atrophy & 4.3\% (n=2,585) & 4.9\% (n=588) & 4.8\% (n=571) & 3.5\% (n=420) & 4.9\% (n=586) & 3.5\% (n=420)\\
Central geographic atrophy & 2.4\% (n=1,455) & 2.9\% (n=344) & 2.9\% (n=347) & 1.8\% (n=216) & 2.7\% (n=325) & 1.8\% (n=223)\\
    \bottomrule
\end{tabular}
\end{table}

\begin{figure}[H]
    \centering
    \frame{\includegraphics[width=\textwidth]{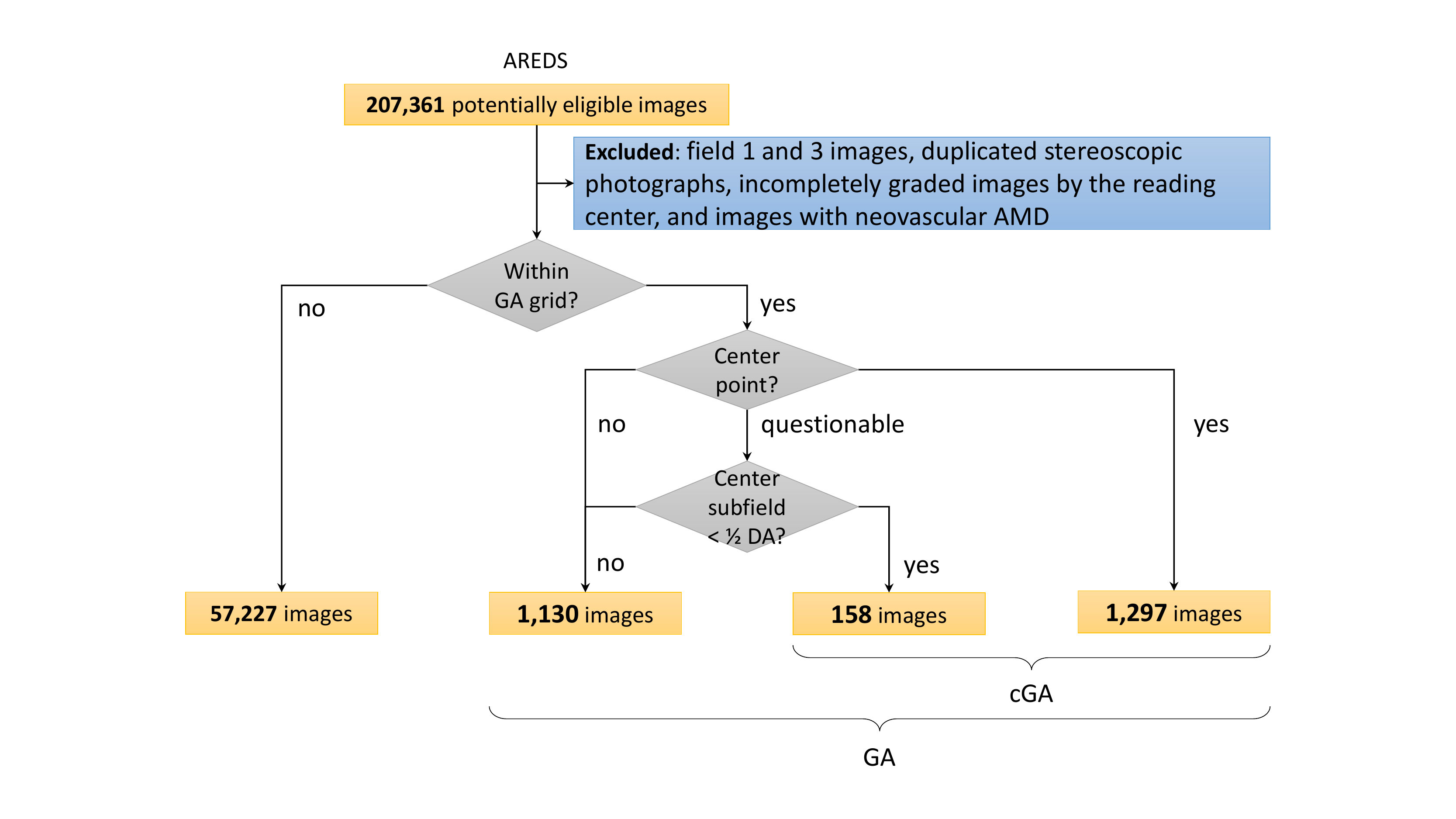}}
    \caption{Distribution of color fundus photographs from the Age-Related Eye Disease Study categorized according to the presence and absence of geographic atrophy (GA) and central geographic atrophy (CGA).}
    \label{fig:Distribution}
\end{figure}

The ground truth labels used for both training and testing were the grades previously assigned to each color fundus photograph by expert human graders at the University of Wisconsin Fundus Photograph Reading Center. (These grading experts did not overlap at all with the 88 retinal specialists described elsewhere). The Reading Center workflow is described in detail in AREDS Report number 69. In brief, a senior grader (grader 1) performed initial grading of the photograph for AMD severity using a 4-step scale (where step 3 includes non-central GA and step 4 includes CGA), and a junior grader (grader 2) performed detailed grading for multiple AMD-specific features (including CGA and non-central GA). In the case of discrepancy between the two graders, a senior investigator at the Reading Center would adjudicate the final grade. All photographs were graded independently, that is, graders were masked to the photographs and grades from previous visits. In addition, a rigorous process of grading quality control was performed at the Reading Center, including the assessment for the inter-grader and intra-grader agreement overall and according to specific AMD features\cite{group2001age}. Inter-grader agreement was substantial for CGA (kappa 0.73). Analyses for potential ``temporal drift'' were conducted by having all graders re-grade in a masked fashion the same group of images annually for the duration of the study. The presence of GA was defined as a sharply demarcated, usually circular, zone of partial or complete depigmentation of the RPE, typically with exposure of underlying large choroidal blood vessels, at least as large as grading circle I-1 (1/8 disc diameter in diameter)\cite{group2001randomized}. If present, GA was categorized as central or non-central according to involvement of the `center point' (Figure~\ref{fig:Distribution}). In this study, images where GA was present alongside signs of neovascular AMD (i.e., image graded positive for both GA and neovascular AMD) were excluded from the dataset.

\subsection*{Models}

The deep learning models for predicting GA and CGA were built using the methods described in DeepSeeNet\cite{peng2018deepseenet}. Specifically, three separate deep learning models were built: (i) one (`GA model') for identifying GA from a population of eyes with no AMD to advanced AMD, irrespective of central involvement, (ii) one (`CGA model') for identifying CGA from a population of eyes with no AMD to advanced AMD, and (iii) one (`centrality detector model') for identifying CGA from a population of eyes with GA (i.e., a model that assesses for central involvement, when GA presence is already known). All three models were binary classification models. As shown in Table~\ref{tab:Characteristics}, there were 2,585 positive instances for the former (GA model) and 1,455 for the latter two (CGA model and centrality detector model). The convolutional neural network architecture used for all models was Inception-v3\cite{szegedy2016rethinking}, which is a state-of-the-art convolutional neural network for image classification. It contains 317 layers, comprising a total of over 21 million weights (learnable parameters) that are subject to training.

The three models were pre-trained using ImageNet, an image database of over 14 million natural images with corresponding labels, using methods described previously\cite{peng2018deepseenet}. (This very large dataset is often used in deep learning to pre-train models. In a process known as transfer learning, pre-training on ImageNet is used to initialize the layers/weights, leading to the recognition of primitive features (e.g., edge detection), prior to subsequent training on the dataset of interest). During subsequent training, for all three models, the input to the model was a color fundus photograph cropped to a square and scaled to 512 pixels. Cropping was done by decreasing the longer dimension of the image to match the shorter dimension (keeping the center point unchanged). We also applied a Gaussian filter to normalize the color balance\cite{grassmann2018deep,bergholm1987edge}. We trained the deep learning models using two commonly used libraries: Keras\cite{chollet2015} and TensorFlow\cite{abadi2016tensorflow}. During the training process, we updated the model parameters using the Adam optimizer (learning rate of 0.0001) for every minibatch of 32 images. Model convergence was measured when the loss on the development set started to increase. The training was stopped after 5 epochs (passes of the entire training set) once the accuracy values no longer increased or started to decrease. All experiments were conducted on a server with 32 Intel Xeon CPUs, using a NVIDIA GeForce GTX 1080 Ti 11Gb GPU for training and testing, with 512Gb available in RAM memory.

\subsection*{Evaluation}

For training and testing, we used the standard machine learning approach of five-fold cross-validation. This consists of subdividing the total data into five equally-sized folds. We used three folds for training, one fold for development (to optimize the hyperparameters and measure the convergence), and the remaining one fold for testing. This process was repeated five times, reserving a different fold each time as the testing set. These settings avoid overfitting the test set, as well as having the test set large enough to be representative. The cross-validation splitting (into the five folds) was performed randomly, at the participant level (i.e., without stratification by disease status). All images from a single participant were present in only one fold. In addition, image augmentation procedures were used, as follows, in order to increase the dataset artificially (i.e., using additional synthetically modified data): (i) rotation, (ii) width shift up to 0.1, (iii) height shift up to 0.1, (iv) horizontal flip, and (v) vertical flip. 
Each model was evaluated against the gold standard Reading Center grades. We obtained the following metrics: overall accuracy, sensitivity, specificity, and precision (with 95\% confidence intervals). Outcomes were also evaluated using receiver operating characteristic (ROC) curves and their corresponding area under the curve (AUC).

The performance of the models was compared with the performance of 88 human retinal specialists (in terms of their pooled gradings). The retinal specialists had graded all color fundus photographs (as part of a qualification survey used to determine AMD severity) at AREDS baseline; the grading was non-overlapping (i.e., each image was graded by one retinal specialist only) and was independent of the Reading Center grading, which was used as the ground truth.

\section*{Results}

\begin{figure}[H]
    \centering
    \frame{\includegraphics[clip, trim=0 2cm 0 2cm ,width=\textwidth]{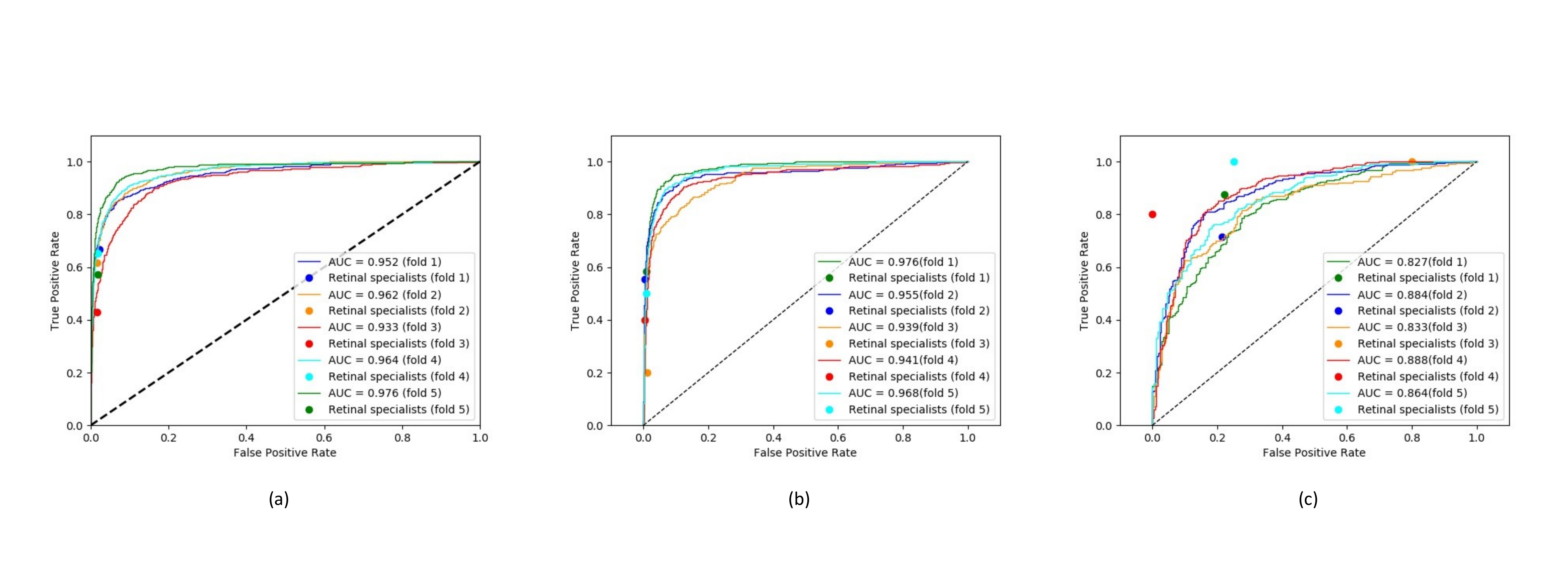}}
    \caption{Receiver operating characteristic curves for (a) the geographic atrophy model, (b) the central geographic atrophy model, and (c) the centrality detector model.}
    \label{fig:roc}
\end{figure}

The ROC curves for the deep learning models are shown in Figure~\ref{fig:roc}. For each model, five ROC curves are displayed on the same graph, with each curve representing one of the folds tested (in the five-fold cross-validation procedure). For comparison, the performance of the human retinal specialists (in terms of their pooled gradings) on the same fold is shown as a single point, such that one point accompanies each of the five curves. For the GA model, the five AUC values were 0.933, 0.952, 0.962, 0.964, and 0.976. For the CGA model, the five AUC values were 0.939, 0.941, 0.955, 0.968, and 0.976. For the centrality detector model, the five AUC values were 0.827, 0.833, 0.864, 0.884, and 0.888. Based on the ROC analysis, the performance levels of the GA model were slightly superior to those of the retinal specialists; those of the CGA model were slightly superior or similar, and those of the centrality detector model were generally inferior.

The accuracy, sensitivity, specificity, and precision of the three models are shown in Table~\ref{tab:Performance}. The GA model demonstrated high accuracy (0.965, 95\% CI 0.959-0.971), attaining sensitivity of 0.692 (0.560-0.825) and specificity of 0.978 (0.970-0.985). The sensitivity was substantially higher and the specificity slightly lower than those of the human retinal specialists. The CGA model also demonstrated high accuracy (0.966, 0.957-0.975), with sensitivity of 0.763 (0.641-0.885) and specificity of 0.971 (0.960-0.982). Again, the sensitivity was substantially higher and the specificity slightly lower than those of the retinal specialists. The centrality detector model had accuracy of 0.762 (0.725-0.799), sensitivity of 0.782 (0.618-0.945), and specificity of 0.729 (0.543-0.916). The accuracy and specificity were slightly higher than those of the retinal specialists, while the sensitivity was substantially lower. For all three models, the precision was higher than that of the retinal specialists.
\begin{table}[ht]
\centering
\scriptsize
\caption{Performance of the three deep learning models: (i) geographic atrophy model (`GA model'), (ii) central geographic atrophy model (`cGA model'), and (iii) centrality detector model.}
\label{tab:Performance}
\begin{tabular}{c@{\hspace{1ex}}c@{\hspace{1ex}}cc@{\hspace{1ex}}cc@{\hspace{1ex}}c}
    \toprule
 & \multicolumn{2}{c}{GA detection} & \multicolumn{2}{c}{cGA detection from all eyes} & \multicolumn{2}{c}{cGA detection from eyes with GA}\\
 \cmidrule(r){2-3} \cmidrule(r){4-5} \cmidrule(r){6-7}
 & GA model & Retinal specialists & cGA model & Retinal specialists & Centrality detector & Retinal specialists\\
  & &  & &  &model & \\
 & (95\% CI) & (95\% CI) & (95\% CI) & (95\% CI) & (95\% CI) & (95\% CI)\\
\midrule
Accuracy & 0.965 (0.959, 0.971) & 0.975 (0.971, 0.980) & 0.966 (0.957, 0.975) & 0.990 (0.987, 0.993) & 0.762 (0.725, 0.799) & 0.735 (0.445, 1.000)\\
Kappa & 0.611 (0.533, 0.689) & 0.438 (0.304, 0.571) & 0.501 (0.451, 0.552) & 0.344 (0.156, 0.532) & 0.513 (0.442, 0.584) & 0.533 (0.176, 0.891)\\
Sensitivity & 0.692 (0.560, 0.825) & 0.588 (0.468, 0.707) & 0.763 (0.641, 0.885) & 0.448 (0.255, 0.641) & 0.782 (0.618, 0.945) & 0.878 (0.722, 1.000)\\
Specificity & 0.978 (0.970, 0.985) & 0.982 (0.978, 0.985) & 0.971 (0.960, 0.982) & 0.993 (0.989, 0.996) & 0.729 (0.543, 0.916) & 0.703 (0.332, 1.000)\\
Precision & 0.584 (0.491, 0.676) & 0.368 (0.230, 0.505) & 0.394 (0.341, 0.448) & 0.296 (0.115, 0.477) & 0.799 (0.710, 0.888) & 0.626 (0.273, 0.979)\\
    \bottomrule
\end{tabular}
\end{table}

\section*{Discussion}

The deep learning model showed relatively robust performance for the detection of GA from a population of eyes with a wide spectrum of disease, from no AMD to advanced AMD. The ROC analysis demonstrated that its performance was non-inferior to that of human retinal specialists. The results of this study highlight the potential utility of deep learning models in identifying GA, based on simple color fundus photographs without additional imaging modalities or other information. In addition, potential utility was demonstrated for deep learning-based detection of CGA in two different scenarios, i.e., from a population of eyes with a wide spectrum of disease and from a population of eyes already known to have GA. Each of these scenarios might be relevant in different clinical or research settings.

To probe further the performance of the GA model that identifies GA presence/absence, particularly in light of lower sensitivity than specificity, we performed an error analysis of false negative cases according to GA area. The hypothesis was that the model performance may have been superior for large GA lesions and inferior for small GA lesions, such that missed cases with small GA lesions might help explain suboptimal model sensitivity. As part of the original Reading Center grading, GA area (within the AREDS grid) was quantified into one of seven categories using the AREDS grading circles\cite{group2001randomized}. We analyzed the proportions of false negatives (GA cases missed by the model) according to the seven area categories. Table~\ref{tab:False} demonstrates that questionable and small GA lesions had a much higher false negative rate (62.9\% for questionable and 78.9\% for area $<$ circle I-2), whereas large GA lesions had a much lower false negative rate (19.8\% for area $\geq$ 2 DA), with evidence of a dose-response effect. Hence, the model's performance had (perhaps understandably) lower sensitivity for small and questionable GA lesions. One contributing factor to this is likely to be the relatively low number of instances during training, since the smaller GA categories accounted for low proportions of the training set.
\begin{table}[ht]
\centering
\caption{False negatives of geographic atrophy detection categorized by the geographic atrophy area within grid.}
\label{tab:False}
\begin{tabular}{lrrr}
    \toprule
\textbf{Area category} & \textbf{Whole test set} & \multicolumn{2}{c}{\textbf{False negatives}}\\
\midrule
Questionable & 275 & 173 & 62.9\%\\
Definite, area $<$ circle I-2 & 38 & 30 & 78.9\%\\
Area $\geq$ I-2 but $<$ O-2 & 125 & 76 & 60.8\%\\
Area $\geq$ O-2 but $<$ \textonehalf~DA & 192 & 90 & 46.9\%\\
Area $\geq$ \textonehalf~DA but $<$1 DA & 297 & 79 & 26.6\%\\
Area $\geq$ 1 DA but $<$2 DA & 403 & 95 & 23.6\%\\
Area $\geq$ 2 DA & 1,255 & 248 & 19.8\%\\
    \bottomrule
\end{tabular}
\end{table}

To probe further the performance of the CGA model that identifies CGA presence/absence, again in light of lower sensitivity than specificity, we performed an error analysis of false negative cases according to the two categories shown in Figure~\ref{fig:Distribution} (i.e., `center point with definite GA' versus `center point with questionable GA but central subfield with definite GA'). As before, we analyzed the proportions of false negatives (CGA cases missed by the model) according to the two categories (Table~\ref{tab:False cga}). As expected, the latter category had a higher false negative rate (31.6\%) than the former category (21.5\%). Hence, the model's performance had (perhaps understandably) lower sensitivity for cases with questionable involvement of the center point.
In future studies, possible approaches to improve the false negative rate in images with small GA lesions might include the use of additional datasets, image augmentation, and generation of synthetic images. In particular, further training on the AREDS2 dataset is likely to improve this aspect of performance, since the dataset includes 1616 eyes with GA (of which 1,118 had GA area $<$0.75 disc areas at first appearance)\cite{keenan2018progression}. In the near future, deep learning might also be able to generate synthetic retinal images with GA (where small lesions could be specified), for improved model training\cite{burlina2019assessment}.
\begin{table}[ht]
\centering
\caption{False negatives of central geographic atrophy detection categorized by involvement of the center point and center subfield.}
\label{tab:False cga}
\begin{tabular}{lrrr}
    \toprule
\textbf{Area category} & \textbf{Whole test set} & \multicolumn{2}{c}{\textbf{False negatives}}\\
\midrule
Definite, center point $<$ I-2 & 1,297 & 279 & 21.5\%\\
Questionable center point, but center subfield area $<$ \textonehalf~disc area & 158 & 50 & 1.6\%\\
    \bottomrule
\end{tabular}
\end{table}

In addition, for the false negative cases, we analyzed the fundus photographs and their accompanying saliency maps (explained in the Supplementary information) for a random sample of 20 images. We graded the photographs qualitatively using pre-specified criteria: (i) image quality, (ii) GA size, (iii) degree of depigmentation in GA lesion(s), and (iv) other factors. Image quality was extremely poor in 5/20 and moderately poor in 10/20. GA size was very small in 4/20 and extremely large (occupying almost all of the photograph) in 2/20. Degree of depigmentation was very mild in 3/20. The most common other factor (in 11/20) was GA lesion features suggestive of prior neovascular AMD (even though the Reading Center grading for these images was negative for neovascular AMD); indeed, examination of previous Reading Center gradings for the same eyes confirmed that, in 10 of these eyes, neovascular AMD had been graded positive at one or more previous study visits. Hence, one or more potential reasons for a false negative grading (very poor image quality, very small GA size, very mild depigmentation, or features of previous neovascular AMD) was found in 19/20 cases. The remaining case had a combination of moderately poor image quality and moderately small GA lesion size. In addition, in 5/20 cases, the saliency maps demonstrated increased signal concentrated in the GA lesion(s), despite the negative prediction.

We consider that these models, and future models arising from them, may be useful in several different scenarios, including assistance in (i) clinical care, (ii) recruitment for clinical trials, (iii) population-based screening for clinical care, and (iv) population-based screening for epidemiological research. Notably, GA has been estimated to affect over 5 million people worldwide\cite{wong2014global,rudnicka2012age}, but it is likely that many of these individuals remain undiagnosed, especially in countries with lower levels of access to retinal specialists. As regards clinical care, the ongoing incorporation of deep learning into radiology and pathology is through augmentation (rather than replacement) of human diagnosis. In ophthalmic care, it might be helpful for an optometrist, general ophthalmologist, or even retinal specialist to have software like this available alongside traditional clinical tools. The software would assist human diagnosis and decision-making, rather than replacing it. For example, it could be used (with settings selected for high sensitivity) by an optometrist, alongside normal examination procedures. Deep learning-assisted GA detection could lead to referral for consultation with a retinal specialist. This approach might also lend itself well to telemedicine approaches, particularly in remote areas. Ultimately, an integrated suite of tools like this might be helpful for augmented human evaluation of a wide range of ophthalmic pathologies, such as AMD, diabetic retinopathy, and glaucoma. However, it would be very important to perform prospective studies to compare the performance of clinicians with and without assistance from deep learning systems.

As to assistance in recruitment for clinical trials, the number of clinical trials of drugs to slow GA enlargement is likely to increase over the next decade. Models like these would enable rapid screening of large numbers of individuals, to identify potential qualifying participants. If treatments that successfully slow GA enlargement are identified and approved, the identification of patients with GA who may stand to benefit from treatment may be aided by these models. Finally, as regards assistance in population-based screening for epidemiological research, accurate data on GA prevalence and characteristics are important for understanding AMD pathophysiology and planning service provision. However, few modern population-based studies have these data readily available. If models like this were used for rapid population-based screening, large quantities of epidemiological data could be produced. For example, the UK Biobank study contains color fundus photographs on 68,544 adult participants (with detailed accompanying data on medical history, lifestyle, physical measures, and blood samples). Applying this model to the dataset would generate a large volume of data on GA prevalence, with powerful links to the medical characteristics of the affected participants. Similar approaches might be possible in the future, e.g. using the Million Veterans Program and others.

There are several limitations to this study. First, the proportions of positive test cases (i.e., images with GA and CGA) were relatively low. This limitation may be reflected in the higher rates of specificity than sensitivity for the first two models. Second, model training and testing were carried out (necessarily) in the context of a high prevalence of AMD, which is a controlled clinical trial rather than a real-world scenario. The ideal setting for evaluating model performance in real-world scenarios (i.e., within a realistic background of retinal diseases and findings) would be a population-based dataset, in which the presence of GA and of other retinal findings had been rigorously graded by human reading center graders. However, the AREDS participants did encounter ocular and systemic disease typical for an elderly population and were not excluded from follow-up for these reasons. Therefore, many AREDS images contain signs of diabetic retinopathy and other pathology (though this was towards the milder end of the disease spectrum, particularly in earlier study years). Hence, model training and testing were performed in the presence of co-existing retinal pathology, perhaps not unlike that expected in population-based studies. Despite this, the GA detection model's specificity was very high, suggesting that the algorithm did not confuse non-AMD lesions with GA. As regards sensitivity, of the sample of 20 false negative cases, we observed no cases where coexisting retinal pathology led to GA being missed.

However, we recommend that the present use case for these models (as currently configured) be limited to research settings, such as in highlighting cases with the potential for enrollment into clinical trials. Before clinical deployment, as with any novel diagnostic aid or test, external validation is required, and should be conducted in the intended use setting(s). The performance of the models for populations like the AREDS, with a very high prevalence of AMD and a lower (but non-zero) prevalence of other retinal diseases and lesions, is expected to be similar to that reported here. We anticipate that the performance may be similar in adult population-based studies, since these also have a relatively low prevalence of non-AMD retinal conditions, though performance might depend on the racial characteristics of the study population. For these reasons, first, we have made freely available the models and code, so that other research groups may perform testing in their own diverse datasets. Second, we are planning external validation studies using well-curated population-based studies from different continents. Following demonstrations of robust external validity and generalizability, these models (or newer models based on them) may be ready for deployment to assist healthcare practitioners in specific clinical settings.

Third, this study is based on one imaging modality only. We used color fundus photographs for GA detection because this imaging modality is the historical standard; color fundus photographs were used in, for example, the AREDS and AREDS2 to define and classify AMD, i.e., to derive classification and prognostic systems including the AREDS 9-step and simplified severity scales\cite{ferris2013clinical,davis2005age}. As such, color fundus photographs are the most highly validated tools for AMD classification and the prediction of disease progression\cite{burlina2018use}. They have the additional advantages of (i) being the closest correlates to biomicroscopy, (ii) consisting of simple datasets (unlike complex 3-dimensional datasets from optical coherence tomography (OCT)), and (iii) (unlike fundus autofluorescence) being widely available and permitting assessment of GA central involvement (without interference from macular pigment in the foveal region).

While OCT may be used increasingly to define GA in the future\cite{sadda2018consensus}, few large standardized datasets currently exist that have followed eyes sequentially for long periods using OCT with detailed accompanying reading center grades and clinical information. Deep learning approaches to OCT datasets hold promise\cite{defauw2018clinically} but sophisticated software is required to process complex 3-dimensional datasets, and automated segmentation of OCT data can be error-prone in disease. Ultimately, we anticipate that deep learning approaches to AMD may be developed using multiple imaging modalities, first using each independently, then combined in a multimodal fashion. In the meantime, the development and validation of deep learning models applied to individual imaging modalities is an important first step.

In conclusion, deep learning demonstrated relatively robust performance for the automated detection of GA. The performance was non-inferior to that of human retinal specialists. In addition, potential utility was observed in the automated identification of CGA. The results of this study may serve as a benchmark for deep learning-based identification of GA from color fundus photographs. We are making the code and pretrained models publicly available at \url{https://github.com/ncbi-nlp/DeepSeeNet}. In this way, the work may help stimulate the future development of automated retinal image analysis tools, not only in the detection of GA, but also in the analysis of GA features including central involvement, multifocality, and area quantification.

\bibliographystyle{vancouver}
\bibliography{references}

\pagebreak

\doublespacing

\section*{Supplement}

\subsection*{Saliency maps}

To visualize important areas in the color fundus images (i.e., those areas that contributed most towards classification), we applied image-specific class saliency maps to assess manually whether the deep learning models were concentrating on image areas that human experts would consider the most appropriate to predict geographic atrophy (GA) or central GA\cite{simonyan2013deep}. The saliency map is widely used to represent the visually dominant location in a given image; it helps highlight areas used by the deep learning algorithm for prediction and can also provide insight into misclassified images. For example, the areas highlighted in the accompanying figure are indeed areas with GA and central GA that are visually apparent in the color fundus images.
\begin{figure}[H]
    \centering
    \frame{\includegraphics[clip, trim=0 0 0 0,width=.6\textwidth]{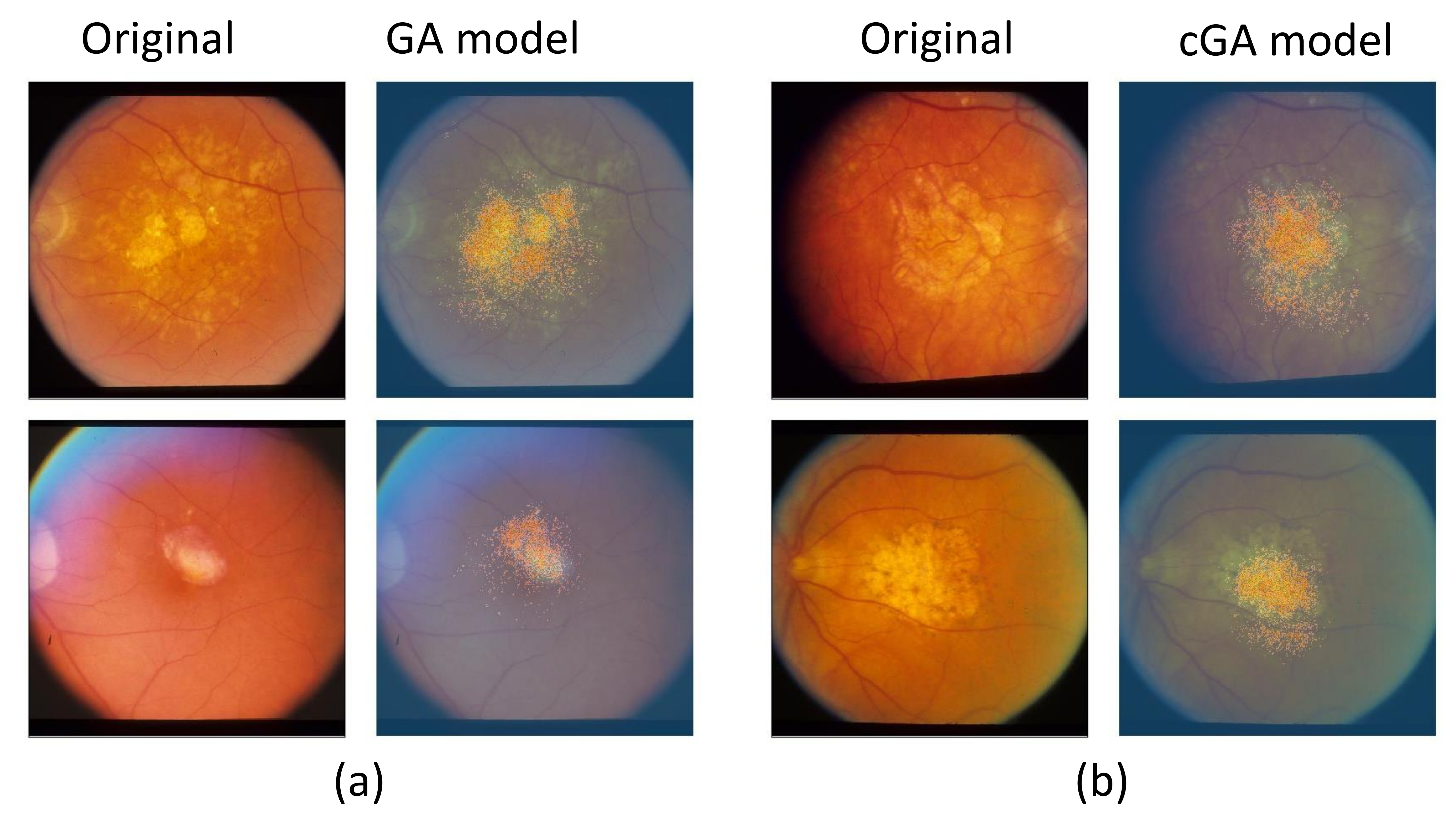}}
    \caption{Representative color fundus photos (first column) side-by-side with overlaid saliency maps (second column), generated by applying keras-vis\cite{raghakotkerasvis} to (a) the geographic atrophy (GA) model and (b) the central geographic atrophy (CGA) model.}
    \label{fig:saliency}
\end{figure}

\end{document}